\documentclass[12pt]{article}
\usepackage{setspace}
\usepackage{hyperref}
\usepackage{geometry}                
\usepackage{dsfont}
\usepackage{amssymb}

\usepackage{amsmath}
\usepackage{cite}
  
\def\toclevel@subsubsubsection{4}
\def\toclevel@paragraph{5}
\def\toclevel@paragraph{6}
\def\l@subsubsubsection{\@dottedtocline{4}{7em}{4em}}
\def\l@paragraph{\@dottedtocline{5}{10em}{5em}}
\def\l@subparagraph{\@dottedtocline{6}{14em}{6em}}

\numberwithin{equation}{section}
\usepackage{dsfont}

\def\mc{\mathcal}
\def\beq{\begin{equation}}
\def\eeq{\end{equation}}
\def\beqn{\begin{eqnarray}}
\def\eeqn{\end{eqnarray}}

\def\dx{{\rm d}^2x}
\def\dy{{\rm d}^2y}

\def\dt{{\rm d}\theta^+}
\def\dtb{{\rm d}\bar{\theta}^+}

\def\Tr{{\rm Tr}}

\def\={\!\!&=&\!}
\def\+{\!\!&+&\!}
\def\-{\!\!&-&\!}

\def\th{\theta^+}
\def\thb{\bar{\theta}^+}
\def\andd{\ \ \ {\rm and} \ \ }
\def\for{\ \ \ {\rm for} \ \ }
\def\with{\ \ \ {\rm with} \ \ }
\def\orr{\ \ \ {\rm or} \ \ }

\newcommand{\ntt}{${\mathcal N}=(2,2)\,\,$}
\newcommand{\nzt}{${\mathcal N}=(0,2)\,\,$}
\newcommand{\ntwo}{${\mathcal N}=2\,\,$}
\newcommand{\none}{${\mathcal N}=1\,\,$}
\newcommand{\tfd}{$2d/4d\,\,$}

\begin{document}

\begin{titlepage}

\begin{flushright}
FTPI-MINN-19/01, UMN-TH-3810/19\\
\end{flushright}

\vspace{1cm}

\begin{center}
\begin{spacing}{1.5}
{  \Large \bf  \boldmath Comments on the NSVZ $\beta$ Functions in Two- dimensional $\mathcal N=(0,2)$ Supersymmetric Models}
\end{spacing}

\vspace{0.8cm}

{\large
Jin Chen$^{a}$ and Mikhail Shifman$^{b,c}$}
\end {center}

\vspace{1mm}

\begin{center}
$^{a}${\it CAS Key Laboratory of Theoretical Physics, Institute of Theoretical Physics, Chinese Academy of Sciences, Beijing 100190, China}\\[1mm]
$^{b}${\it  Department of Physics, University of Minnesota,
Minneapolis, MN 55455, USA}\\[1mm]
$^c${\it  William I. Fine Theoretical Physics Institute,
University of Minnesota,
Minneapolis, MN 55455, USA}\\[1mm]

\end{center}

\vspace{0.5cm}

\begin{center}
{\large\bf Abstract}
\end{center}

The NSVZ $\beta$ functions in two-dimensional $\mathcal N=(0,2)$ supersymmetric models are revisited. We construct and discuss a broad class of such models using the gauge formulation. All of them represent direct analogs of four-dimensional ${\mathcal N} =1$ Yang-Mills theories and are free of anomalies. Following the same line of reasoning as in four dimensions we distinguish between the holomorphic and canonical coupling constants. This allows us to derive the {\em exact} two-dimensional $\beta$ functions in all models from the above class. We then compare our results with a few examples 
which have been studied previously.

\hspace{0.3cm}

\end{titlepage}

\tableofcontents

\newpage


\section{Introduction and conclusion}
\label{intro}

The $2d/4d$ parallels are known and were used since the time of Polyakov who found asymptotic freedom (AF) in $2d$ non-linear sigma models  \cite{Polyakov}, in analogy with AF in $4d$ Yang-Mills theories \cite{Gross, Politzer}. In the last three decades, $2d/4d$ correspondence acquired a much deeper meaning  by virtue of supersymmetry. Much of non-perturbative dynamics in both $2d/4d$ supersymmetric gauge theories has been thoroughly understood and found to correspond to each other. By the ``$2d/4d$ correspondence" we mean here the cases in which either some of  \tfd physics contents are exactly the same, e.g. the Alday-Gaiotto-Tachikawa (AGT) correspondence \cite{AGT}, or the dynamical behaviors in 2d and 4d coincide, for instance, the BPS spectra, certain correlation functions, dualities, etc., are identical \cite{bps, CCSV, tt, CJ, Seiberg, Gukov}. Among these phenomena, an instructive  example is provided by non-Abelian BPS vortex strings \cite{Tong, Yung, Yung2}, both in $4d$ \ntwo and \none gauge theories, whose low-energy dynamics are captured by $2d$ \ntt and heterotic \nzt sigma models respectively \cite{Tong1, SY1, Tong2, Tong3, SY2}. The above vortex strings present a ``bridge'' between $4d$ and $2d$  physics providing a quantitative explanation why the $2d$ dynamics are in correspondence with the dynamics in its $4d$ progenitor.  This correspondence was established in a wide class of theories both from $2d$ and $4d$ directions, perturbatively and non-perturbatively \cite{SY3, SY4, SY5, CS1, CS2, CS3, CCSV}. 

The goal of this paper is to derive NSVZ-like $\beta$ functions \cite{NSVZ2, NSVZ3, NSVZ4, NSVZ, Shifman} in general two-dimensional \nzt supersymmetric gauge theories adding new evidence for the $2d/4d$ correspondence. A number of $2d$ analogs of the NSVZ $\beta$ functions were obtained in the past via both perturbative methods and instanton calculus in the \nzt $\mathds{CP}^1$ model \cite{CS3} and in a large class of heterotically deformed non-linear sigma models (NLSMs) which are
deformations of their \ntt cousins \cite{CCSV}. Here we focus on another general class of \nzt gauged linear sigma models (GLSMs) and obtain 
 the general form of the corresponding $\beta$ functions. They have the same structure as the NSVZ $\beta$ function in $4d$. In those cases where comparison with the previous results is possible our newly derived GLSM $\beta$ functions are  identical to those of NLSMs. This is not surprising since the NLSMs studied previously can be embedded in GLSMs. 

We want to emphasize not only the ubiquity of \tfd correspondence, but also the conspiracy of methodologies applicable to both $2d$ and $4d$ theories. Historically, $2d$ sigma models were considered as simplified toy models useful for understanding real world physics in $4d$. Instead, in this paper, we follow the opposite direction, from $4d$ to $2d$, establishing and using the $2d$ analog of the Konishi anomaly \cite{Konishi} and scaling anomalies  in $2d$ \nzt gauge theories, \emph{\`a la} Arkani-Hamed and Murayama in $4d$ $\mathcal N=1$ case \cite{AHM}. This observation helps us relate holomorphic
coupling constants to canonic ones in $2d$ GLSMs thus trivializing derivation of their $\beta$ functions.  The general master formula obtained in this paper is  
\beq
\beta(g^2)=-\frac{g^4}{4\pi}\frac{\sum_i q_i+\frac{1}{2}\sum_a{\tilde q_a}\gamma_a}{1-\frac{\sum_i q_i}{8\pi} g^2}\,,
\label{master}
\eeq 
in the case of $2d$ \nzt gauge theories with a single FI coupling 
$$\xi\equiv\frac{2}{g^2}\,,$$ 
where $q_i$'s are the $U(1)$ gauge charges of the bosonic matter
fields, $\tilde q_a$ and $\gamma_a$'s are the $U(1)$ gauge charges and anomalous dimensions of the fermionic matter fields.\\ 

The paper is organized as follows: We will briefly review the building blocks of $2d$ \nzt supersymmetric GLSMs in section \ref{2} and a non-renormalization theorem for the FI coupling constants in section \ref{3}. We then explain the difference between holomorphic and canonical coupling constants both from the perspectives of the Konishi anomaly and the scaling anomalies of matter fields, and derive the master equation (\ref{master}) in section \ref{4}. Finally, we apply the formula in several examples.\\

\section{Two-dimensional \nzt GLSMs}
\label{2}

The \nzt superspace is parametrized by $2d$ bosonic spacetime $$x^{\pm\pm}\equiv x^0\pm x^1$$ and their \nzt fermionic partners $\theta^+$ and $\bar{\theta}^+\,.$ The supercharges are defined in terms of these coordinates  as follows:
\beqn
&&Q_+\equiv\frac{\partial}{\partial\theta^+}+i\bar{\theta}^+\partial_{++}\,,\nonumber\\[2mm]
&&\bar{Q}_+\equiv-\frac{\partial}{\partial\bar{\theta}^+}-i\theta^+\partial_{++}\,,
\eeqn
where $$\partial_{++}\equiv 2\partial_{x^{++}},\ \partial_{--}\equiv 2\partial_{x^{--}}\,.$$ Accordingly, the superderivatives are given by
\beqn
&&D_+\equiv\frac{\partial}{\partial\theta^+}-i\bar{\theta}^+\partial_{++}\,,\nonumber\\[2mm]
&&\bar{D}_+\equiv-\frac{\partial}{\partial\bar{\theta}^+}+i\theta^+\partial_{++}\,,
\eeqn
which satisfy the conditions $$D_+^2=\bar{D}_+^2=0\,, \qquad \{D_+,\bar{D}_+\}=2i\partial_{++}\,.$$ With this notation, it is not difficult to build three types of supermultiplets to construct \nzt GLSMs \cite{Witten, Tong1}.\\

\noindent{\bf Gauge multiplets}:

\vspace{2mm}

\noindent The \nzt gauge multiplet $U_{--}=(A_{--},\,\lambda_-,\,\bar{\lambda}_-,\,D)$ is real and adjoint-valued
\beq
U_{--}=A_{--}-2i\theta^+\bar{\lambda}_- -2i\bar{\theta}^+\lambda_- +2\theta^+\bar{\theta}^+D
\eeq
in superfield formalism. Here $$A_{--}\equiv A_0-A_1\,,\qquad A_{++}\equiv A_0+A_1$$ are the $2d$ gauge fields, $\lambda_-$ and $\bar{\lambda}_-$ are the gaugino fields, and the real field $D$ is auxiliary. The field $A_{++}$  is an \nzt singlet. 

Next, we can promote superderivatives to be covariant, namely
\beqn
&&\mathcal D_+\equiv\frac{\partial}{\partial\theta^+}-i\bar{\theta}^+\nabla_{++}\equiv\frac{\partial}{\partial\theta^+}-i\bar{\theta}^+(\partial_{++}-iA_{++})\,,\nonumber\\[2mm]
&&\bar{\mathcal D}_+\equiv-\frac{\partial}{\partial\bar{\theta}^+}+i\theta^+\nabla_{++}\equiv-\frac{\partial}{\partial\bar{\theta}^+}+i\theta^+(\partial_{++}-iA_{++})\,,\nonumber\\\nonumber\\[1mm]
&&\mathcal D_{--}\equiv\partial_{--}-iU_{--}=\nabla_{--}-2\theta^+\bar{\lambda}_--2\bar{\theta}^+\lambda_--2i\theta^+\bar{\theta}^+D\,.
\label{superder}
\eeqn
The superfield strength of the gauge multiplet is given by
\beq
\Upsilon_-=\left[\bar{\mc D}_+,\mc D_{--}\right]=-2\left(\lambda_--i\theta^+(D-iB)-i\theta^+\bar{\theta}^+\mc D_{++}\lambda_-\right)\,,
\label{sstrength}
\eeq
where 
\beq
B=\partial_0A_1-\partial_1 A_0-i\left[A_0,A_1\right]
\label{bee}
\eeq
 is the field strength of the $A_\mu$ field. 
The conjugated superfield $\bar\Upsilon_-$ is defined accordingly.
The action of the gauge multiplet is as follows:
\beq
S_{\rm gauge}=\frac{1}{8e^2}\Tr\int\dx\,\dt\dtb\,\overline{\Upsilon}_-\Upsilon_-=\frac{1}{e^2}\Tr\int\dx\left(\frac{1}{2}B^2+i\bar{\lambda}_-\nabla_{++}\lambda_-+\frac{1}{2}D^2\right)\,.
\eeq\\
Here $e^2$ is the gauge coupling. The corresponding NLSM can be obtained in the limit $e^2\to \infty$.\\

\noindent{\bf Chiral multiplets:}

\vspace{2mm}

\noindent The \nzt chiral multiplet $\Phi^i=(\phi^i,\, \psi_+^{i})$ satisfies the usual chiral constraint 
\beq
\bar{\mc D}_+\Phi^i=0\,.
\eeq
In the superfield formalism it is written as
\beq
\Phi^i=\phi^i+\sqrt{2}\theta^+\psi_{+}^{i}-i\th\thb\nabla_{++}\phi^i\,,
\eeq
where $$\nabla_{\mu}\phi^i=(\partial_{\mu}-iq_i A_{\mu})\phi^i\,.$$ Moreover, $q_i$ is the charge of the field $\Phi^i$ with respect to the $U(1)$ gauge field. The action of the chiral multiplets can be written as
\beqn
S_{\rm chiral}\=-\frac{i}{2}\int\dx\,\dt\dtb \sum_i\overline\Phi_i\mc D_{--}\Phi^i\nonumber\\[1mm]
\=\int\dx\sum_i\left(-\left|\nabla_\mu\phi^i\right|^2+i\bar\psi_{+\,i}\nabla_{--}\psi_+^{i}-\sqrt{2}iq_i\bar{\phi}_i\lambda_-\psi_+^i+\sqrt{2}q_i\bar{\psi}_{+\,i}\bar{\lambda}_-\phi^i+q_i\bar\phi_iD\phi^i\right).\nonumber\\
\label{S1}
\eeqn

\noindent{\bf Fermi multiplets:}

\vspace{2mm}

\noindent  Another important matter superfield consists of a fermion $\chi^a_-$ and an auxiliary field $G^a$,
\beq
\left(\chi^a_-,\, G^a,\right) \in \Gamma^a_-\,.
\eeq
It is not necessary chiral, but, instead, satisfies the constraint
\beq
\bar{\mc D}_+\Gamma^a_-=\sqrt{2}E^a(\Phi)\,,
\eeq
where $E(\Phi)$ is an arbitrary holomorphic function with respect to chiral boson fields $\Phi$'s. In the superfield formalism, it can be expanded as
\beq
\Gamma^a_-=\chi^a_--\sqrt{2}\theta^+G^a-i\th\thb\nabla_{++}\chi_--\sqrt{2}\thb E^a(\Phi)\,.
\eeq
The action for the fermi multiplet reduces to
\beqn
S_{\rm fermi}\=-\frac{1}{2}\int\dx\,\dt\dtb\,\sum_a \overline{\Gamma}_{-\,a}\Gamma_-^a\nonumber\\
\=\int\dx\sum_{a,\,i}\left(i\bar{\chi}_{-\,a}\nabla_{++}\chi_-^a+\left|G^a\right|^2-\left|E^a(\phi)\right|^2-\bar\chi_{-\,a}\frac{\partial E^a}{\partial\phi^i}\psi^i_++{\rm h.c.}\right).\nonumber\\
\label{S2}
\eeqn
Note that the gauge field strength $\Upsilon_-$ is a particular case of the fermi multiplets in the adjoint representation of the gauge group, satisfying
\beq
\bar{\mc D}_+\Upsilon_-=0\,.
\eeq

\noindent{\bf Superpotentials:}

\vspace{2mm}

\noindent Last but not least, we need to introduce superpotentials $J_a(\Phi)$ as holomorphic functions of chiral superfields, whose action reduces to a half of the superspace (accompanied by fermi multiplets $\Gamma^a_-$),
\beqn
S_J \= -\frac{1}{\sqrt{2}}\sum_a\int\dx\,\dt\,\Gamma^a_- J_a+{\rm H.c.}\nonumber\\[2mm]
\=\sum_a\int\dx\, G^a J^a(\phi)+\sum_i\chi_{-\,a}\frac{\partial J^a}{\partial\phi^i}\psi^i_++{\rm H.c.}\,.
\eeqn
Of the utmost interest is the Fayet-Iliopoulos (FI) term as a superpotential given by the gauge field strength, if it admits $U(1)$ factors,
\label{Stau}
\beq
S_\tau=\frac{1}{4}\Tr\int\dx\,\dt\,\tau\,\Upsilon_{-}|_{\thb=0}+{\rm h.c.}=\Tr\int\dx\left(-\xi D+\frac{\theta}{2\pi}B\right)\,,
\label{fit}
\eeq
where for simplicity we only consider theories with a single FI term, and 
\beq
\tau=\frac{\theta}{2\pi}+i\xi
\eeq
is the complexitied FI coupling constant.\\

\noindent{\bf GLSM action:}

\vspace{2mm}

\noindent Overall we assemble all the above ingredients and arrive at the action of \nzt supersymmetric GLSM,
\beq
S=S_{\rm gauge}+S_{\rm chiral}+S_{\rm fermi}+S_{\tau}+{\rm H.c.}\,.
\label{S}
\eeq
Here and below, without loss of generality, we will consider theories in which the superpotentials are limited to  FI terms. Importantly, for such theories to be consistent at the quantum level (i.e. free of internal anomalies), we need to impose constraints on the representations of 
the chiral and fermi multiplets to get rid of the gauge anomalies, see also in \cite{CCSV2},
\beqn
&&U(1)\ \ {\rm gauge}:\ \ \ \ \ \sum_i q_i^2=\sum_a \tilde q_a^2\,,\nonumber\\
&&{\rm non\!\!-\!\!Abelian\ \ gauge}:\ \ \ \ \ \sum_i t_2(i)=t_2(A)+\sum_a t_2(a)\,,
\eeqn
where 
$q_i$ and $\tilde q_a$ are $U(1)$ gauge charges of chiral and fermi multiplets, $t_2$ is the dual Coxeter number, and ``$i$", ``$a$" and ``$A$" denote the Reps. of chiral, fermi and gauge multiplets.\\

\section{A non-renormalization theorem for the holomorphic coupling $\tau$}
\label{3}

In $2d$ gauge theories, the gauge coupling $e$ has  dimension of mass, and is thus superrenormalizable. For energy scale $\mu\ll e$, the gauge multiplets will be non-dynamical and we arrive at NLSMs. Therefore the only sensible parameter in the theory is its FI coupling constant $\tau$, which is marginal and runs at the quantum level. In much the same way as with the gauge couplings in $4d$ \none gauge theories, the $2d$ FI parameter $\tau$, as the coupling of the \nzt superpotential, is subject to a non-renormalization theorem and receives at {\em most one-loop} correction (see e.g. \cite{NSVZ}) . We will follow \cite{NSVZ,AHM} in reviewing the relevant argument.\\

\noindent From eq.(\ref{S}), we see that the action $S$ depends on $\tau$ holomorphically. It is convenient to use the notation
\beq
2\pi i\tau=-2\pi\xi+i\theta\equiv-\frac{4\pi}{g^2}+i\theta\,.
\eeq
Let us ask ourselves: when we change the cutoff from $M_0$ to $\mu$, how the coupling $2\pi i\tau(\mu)$ (in the Wilsonian sense) changes to keep the low-energy physics intact. To answer this question, let us examine an ansatz
\beq
2\pi i\tau\left(\mu\right)=2\pi i\tau\left(M_0\right)+f\left(2\pi i\tau(M_0),\log\frac{M_0}{\mu}\right)\,.
\eeq
It is worth noting that a $2\pi$ shift of the $\theta$ angle leads no change of physics, therefore at most,
\beq
f\left(2\pi i\tau(M_0),\log\frac{M_0}{\mu}\right)\longrightarrow f\left(2\pi i\tau(M_0),\log\frac{M_0}{\mu}\right)+2\pi i\, F\left(\log\frac{M_0}{\mu}\right)\,,\ \ {\rm for}\ \ \theta\rightarrow\theta+2\pi\,,
\eeq
where function $F\left(\log\frac{M_0}{\mu}\right)$ can only take integer values. Furthermore because $F(0)=0$, by continuity we conclude that function $f$ is periodic respect to the $\theta$ angle. Therefore the $\beta$ function for $2\pi i\tau$,
\beq
\beta(2\pi i\tau)=\mu\frac{\partial}{\partial\mu}\left(2\pi i\tau(\mu)\right)=\mu\frac{\partial f}{\partial\mu}\,,
\eeq
is periodic with respect to $\theta$ and admits a Fourier expansion,
\beq
\beta(2\pi i\tau)=\sum_{n\geq 0}b_n\,e^{2\pi i n\tau}\,.
\label{expa}
\eeq
It is clear that in perturbation theory we can only have non-negative integer values of $n$ appearing in the expansion
(\ref{expa}). Also, in the perturbative regime we at most have $b_0$ nonzero, i.e.
\beq
\beta(2\pi i\tau)=b_0\,,
\label{beta}
\eeq
It perturbation theory it is obvious that all $b_n$'s with $n= 1,2,3,...$ vanish.
Hence the non-renormalization theorem of the absence of higher loops is proven for the holomorphic coupling. 

Non-perturbatively, one needs to apply the anomalous $R$-symmetry of \nzt, which guarantees that the $\theta$ angle receives no quantum corrections at all. Consequently $\beta(2\pi i\tau)$ is independent of ${\rm Im}(2\pi i\tau)$, \emph{and, simulateneously is holomorphic} in $2\pi i\tau$. It implies that $\beta(2\pi i\tau)$ can only be  a constant, i.e. eq.(\ref{beta}) holds both perturbatively and non-perturbatively.

Before proceeding to the discussion of the canonical coupling $\tau_{\rm c}$ in next sections, let us first calculate $b_0$ that would be used latter. It can be easily obtained by inspecting the $D$ term of the action (\ref{S}),
\beq
S_D=\int\dx\left(\frac{1}{2e^2}D^2-\xi D+\sum_i q_i\bar\phi_i D\phi^i\right)\,.
\label{D}
\eeq
From (\ref{D}) we see that the real part of $\tau$ receives a tadpole one-loop correction.\footnote{As in the $4d$ case, the tadpole correction appears if and only if $\sum_i q_i \neq 0$.} The tadpole graph emerges through contracting $\phi$ and $\bar\phi$. As a result,
\beq
\xi(\mu)=\xi(M_0)-\frac{\sum_i q_i}{2\pi}\log\left(\frac{M_0}{\mu}\right)\,,
\eeq
which implies, in turn, that
\beq
\beta(\xi)=\frac{\sum_i q_i}{2\pi}\,,\ \ {\rm or,\ \ say,}\ \ \beta(g^2)=-\frac{\sum_i q_i}{4\pi}\,,
\label{1loop}
\eeq
and
$$  b_0=-\sum_i q_i\,.$$

\section{From the holomorphic to canonic coupling}
\label{4}

As known from \cite{NSVZ}, all higher order loops in the gauge coupling renormalization appear
in passing from the holomorphic to canonic coupling from the $Z$ factors of the matter fields (which are converted into the anomalous dimensions
in the $\beta$ functions). To see how this happens we must convert the kinetic terms of the matter fields into (\ref{fit}) by virtue of anomalies. In other words, we must take into account a subtle difference between the Wilsonian Lagrangian and 1PI irreducible functional (see \cite{NSVZ,NSVZ2, NSVZ3, NSVZ4}).

Below we will
discuss two alternative (but related) derivations, through the Konishi anomaly \cite{Konishi} and through the scale anomaly \cite{AHM}. 

\subsection{The Konishi anomaly in ${\mathcal N}=(0,2)$ GLSM}
\label{kon}

It is not difficult to establish the $2d$ analog of the Konishi anomaly. To this end, as an example, we will consider the operator 
$\sum_a \overline{\Gamma}_{-\,a}\Gamma_-^a$ appearing in (\ref{S2}) (assuming that $E^a=0$).
Classically, the equation of motion for this operator is
\beq
{\mc D}_+\left(\mathcal \sum_a \overline{\Gamma}_{-\,a}\Gamma_-^a\right) =0\,.
\eeq
This follows, e.g. from inspection of the $\bar\theta^+$ component. However, at the quantum level this particular component
contains a well-known anomaly in the derivative of the $\chi_-$ current, see more details in appendix \ref{B} and also \cite{CCSV1},  analogous to the triangle anomaly in the axial current in $4d$,\footnote{The triangle anomalous graph in four dimensions is replaced in two dimensions by a diangle graph. That's why the right-hand side in (\ref{miril}) is linear in $\tilde{q}_a$.}
\beq
\left. \partial_{++} \left(\sum_a\bar\chi_{-\,a}\chi_-^a\right)= \sum_a\, \frac{\tilde{q}_a}{2\pi} B\right|_{U(1)}
\label{miril}
\eeq
where $B$ is defined in (\ref{bee}). Note that the relative coefficient between $D$ and $B$ in (\ref{sstrength}) is rigidly fixed by \nzt supersymmetries. Needless to say, that
the full derivative in the $U(1)$ part does not appear in the action classically (it can be dropped). 
However, at the quantum level we can establish the following relations (after evolving the action  from $M_0$ down to $\mu$),
\beqn
\Delta{\mathcal L}_\Gamma (\mu ) &&-\frac 12 Z_{\rm fermi} \int d\bar\theta^+\, d\theta^+ \Big(\overline{\Gamma}_{-\,a}\Gamma_-^a\Big)  = -\frac{Z_{\rm fermi}}{2}\int 
d\bar\theta^+\, {\mathcal D}_+ \Big(\overline{\Gamma}_{-\,a}\Gamma_-^a\Big)\nonumber\\[2mm]
&& =i\frac{Z_{\rm fermi}}{2}\partial_{++}\left(\sum_a\bar\chi_{-\,a}\chi_-^a\right)=\left.iZ_{\rm fermi}\sum_a\frac{\tilde q_a}{{4\pi}} B\right|_{U(1)}\nonumber\\[2mm]
&&=\left.iZ_{\rm fermi}\sum_a\frac{\tilde{q}_a}{8\pi}\left(\int\dt\Upsilon_-+\int d\bar\theta^+\, \bar{\Upsilon}_-\right)\right|_{U(1)}\,,
\eeqn
where in the last step, we uplifted the equation to the level of superspace, cf. (\ref{fit}). The $\Upsilon_-$ part  gives the evolution of the wave function renormalization of fermion $\Gamma_-^a$ to the FI-coupling constant $\tau$, see also eq.(\ref{upsilon}). Adding the one-loop tadpole graph and differentiating over $\mu/\partial\mu$ we arrive at the $\tilde{q}_a\gamma_a$ term in (\ref{master}).

\subsection{Scaling anomalies}
\label{4.2}

Now we would like to discuss the $2d$ \nzt $\beta$ function along the the lines of \cite{AHM}. It is true that the holomorphic $\tau$ only receives one-loop correction, however, because of the normalization point running down from $M_0$ to $\mu$, the kinetic terms of the matter fields will receive a wave function renormalization,
\beq
\sum_i\overline\Phi_i\mc D_{--}\Phi^i\longrightarrow \sum_i Z_i(\mu)\,\overline\Phi_i\mc D_{--}\Phi^i\,,\ \ \ \sum_a \overline{\Gamma}_{-\,a}\Gamma^a_-\longrightarrow \sum_a Z_a(\mu)\,\overline{\Gamma}_{-\,a}\Gamma^a_-\,,
\eeq
see section \ref{kon} for $\overline{\Gamma}_{-\,a}\Gamma_-^a$.

To keep all matter fields canonically normalized, we need to change field variables, i.e. redefine
\beq
\Phi^i\equiv\frac{1}{\sqrt{Z_i(\mu)}}\Phi^{i\,\prime}\,,\ \ \ \Gamma^a_-\equiv\frac{1}{\sqrt{Z_a(\mu)}}\Gamma^{a\,\prime}_-\,.
\eeq
However, such rescaling will result in anomalous Jacobians from the functional measure. Formally we have
\beqn
&&\left[d\Phi^i\right]=\left[d\left(\frac{1}{\sqrt{Z_i(\mu)}}\Phi^{i\,\prime}\right)\right]={\rm sDet}\left(\frac{1}{\sqrt{Z_i(\mu)}}\right)\left[d\Phi^{i\,\prime}\right]=\left[d\Phi^{i\,\prime}\right]e^{-\frac{1}{2}\log Z_i(\mu)\,{\rm sTr}_{\Phi^i}\mathds 1}\,,\nonumber\\[2mm]
&&\left[d\Gamma^a_-\right]=\left[d\left(\frac{1}{\sqrt{Z_a(\mu)}}\Gamma^{a\,\prime}_-\right)\right]={\rm sDet}\left(\frac{1}{\sqrt{Z_a(\mu)}}\right)\left[d\Gamma^{a\,\prime}_-\right]=\left[d\Gamma^{a\,\prime}_-\right]e^{-\frac{1}{2}\log Z_a(\mu)\,{\rm sTr}_{\Gamma^a_-}\mathds 1}\,,\nonumber\\
\label{Jacob}
\eeqn
where ``sDet" and ``sTr" denote the super-determinant and super-trace, respectively. The super-trace is superficially vanishing due to supersymmetries. Nevertheless, in a non-trivial gauge field background, we can show that they give rise to terms proportional to the $U(1)$ field strength $\Upsilon_-$. More specifically,
\beq
{\rm sTr}_{\Phi^i}\mathds 1=-i\frac{q_i}{8\pi}\int\dx\,\dt\,\Upsilon_-|_{\thb=0}\,,\ \ \ {\rm and}\ \ \ {\rm sTr}_{\Gamma^a_-}\mathds 1=i\frac{\tilde q_a}{8\pi}\int\dx\,\dt\,\Upsilon_-|_{\thb=0}\,.
\label{upsilon}
\eeq
The derivation of this formula is presented in appendix \ref{B}. Therefore, the holomorphic $\tau$ will receive \emph{non-holomorphic} corrections from wave function renormalizations,
\beq
\tau\longrightarrow\tau_c=\tau+\sum_i i\frac{q_i}{4\pi}\log Z_i(\mu)-\sum_a i\frac{\tilde q_a}{4\pi}\log Z_a(\mu)\,.
\label{tc}
\eeq
The anomalous dimensions of $\Phi^i$ and $\Gamma^a_-$ are given by
\beq
\gamma_i=-\mu\frac{\partial}{\partial\mu}\log Z_i(\mu)\,,\ \ {\rm and}\ \ \gamma_a=-\mu\frac{\partial}{\partial\mu}\log Z_a(\mu)\,
\eeq
and they are non-holomorphic. This statement is in one-to-one correspondence with the NSVZ $\beta$ function in four dimensions.

Differentiating $\log\mu$ on both sides of eq.(\ref{tc}) and using eq.(\ref{1loop}), we have
\beq
\beta(\tau_c)=i\left(\frac{\sum_i q_i}{2\pi}-\sum_i\frac{q_i}{4\pi}\gamma_i+\sum_a\frac{q_a}{4\pi}\gamma_a\right)\,.
\label{beta2}
\eeq
In terms of coupling constant 
\beq
{\rm Im}(\tau_c)=\xi_c\equiv\frac{2}{g_c^2}\,
\eeq
we have
\beq
\beta(g_c^2)=-\frac{g_c^4}{4\pi}\left(\sum_i q_i-\frac{1}{2}\sum_i q_i\gamma_i+\frac{1}{2}\sum_a \tilde q_a\gamma_a\right)\,.
\eeq
Furthermore,  from eq.(\ref{D}), the $\beta$ function of $g_c^2$, or say, $\xi$, is nothing other than the wave function renormalization of chiral multiplets, i.e.
\beq
\gamma_i=\frac{\beta(g_c^2)}{g_c^2}\,.
\label{betagamma}
\eeq 
Using it, we arrive at the master formula,
\beq
\beta(g_c^2)=-\frac{g_c^4}{4\pi}\frac{\sum_i q_i+\frac{1}{2}\sum_a{\tilde q_a}\gamma_a}{1-\frac{\sum_i q_i}{8\pi} g_c^2}\,.
\eeq 

\vspace{2mm}
\noindent
Remark: The gauge multiplets have no contribution to the $\beta$ function, because $\tau_c$ is associated with the $U(1)$ factor gauge group, with respect to which the gauge multiplet is $U(1)$ neutral.\\

\section{Examples}
\label{5}

In this section, we will apply eq.(\ref{master}) in various examples.

\subsection{${\mathcal N}=(2,2)\,\,\mathds{CP}^{N-1}$ model}

For \ntt supersymmetries, the \nzt chiral and fermi multiplets are combined to an \ntt chiral multiplet. We have
\beq
q_i=\tilde q_a\,,\ \ {\rm and}\ \ Z_i=Z_a\,,\ \ {\rm for}\ \ i=a=1,2,\dots
\eeq
Therefore the holomorphic $\tau$ and canonical $\tau_c$ coincide, and the $\beta$-function terminates at one-loop, in terms of $g_c^2$,\footnote{Exactly the same occurrs in 4d Yang-Mills \cite{NSVZ, Shifman}.}
\beq
\beta(g_c^2)=-\frac{\sum_i q_i}{4\pi}g^4_c
\eeq\\
Especially, for a $U(1)$ gauge theory with all $q_i=1$, we have the standard \ntt $\mathds{CP}^{N-1}$ sigma model, and its $\beta$-function is
\beq
\beta(g_c^2)=-\frac{N}{4\pi}g^4_c\,.
\eeq

\subsection{\nzt $\mathds{CP}^{N-1}$ model}
\label{5.3}
We can deform the previous \ntt $\mathds{CP}^{N-1}$ model by deleting part of \ntt $U(1)$ field strength, considered in \cite{Tong2}. In the language \nzt supersymmetries, the \ntt $U(1)$ field strength $\Sigma_{(2,2)}$ can be decomposed as,
\beq
\Sigma_{(2,2)}=\Sigma_{(0,2)}\oplus \Upsilon_-\,,
\eeq
where the $\Sigma_{(0,2)}$ is a \nzt chiral superfield and $\Upsilon_-$ is the \nzt fermi multiplet as the field strength of $U(1)$ gauge multiplet. \ntt chiral multiplet $\Phi_{(2,2)}^i$ also admits a decomposition as
\beq
\Phi_{(2,2)}^i=\Phi^i\oplus \Gamma_-^i\,,
\eeq
and the \nzt fermi multiplet $\Gamma_-^i$ satisfy the constraint
\beq
\bar{\mathcal D}_+\Gamma_-^i\propto \Sigma_{(0,2)}\Phi^i\,.
{\label{fc22}}
\eeq
Now, if we delete $\Sigma_{(0,2)}$, the deformed theory will have only \nzt supersymmetry, and the fermi multiplets satisfy
\beq
\bar{\mc D}_+\Gamma_-^i=0\,.
\label{fc}
\eeq
Its $\beta$ function turns out to be
\beq
\beta(g_c^2)=-\frac{N g_c^4}{4\pi}\frac{1+\frac{1}{2}\gamma}{1-\frac{N}{8\pi} g_c^2}\,,
\eeq
where $\gamma$ denotes the anomalous dimension of Fermi multiplet $\Gamma_-^i$. We want to further comment that, in \cite{CS3}, the authors also considered a type of deformed \nzt $\mathds{CP}^1$ model at the level of NLSM, which is different from ours. However, we do see that the $\beta$ functions of the two models are similar. To compare the difference between our model and that in \cite{CS3}, we discuss its non-linear formalism in appendix \ref{A}.\\

\subsection{Heterotically deformed \nzt $\mathds{CP}^{N-1}$ model}

We can also consider a further deformation from the \nzt $\mathbb{CP}^{N-1}$ model discussed above, by adding an additional \emph{gauge singlet} \nzt fermi multiplet,
\beq
\Omega_-=\eta_--\sqrt{2}\theta^+H-i\theta^+\bar\theta^+\partial_{++}\eta_-\,,
\eeq
to the \nzt model, with the corresponding deformed term in the action,
\beq
\mc S_{\Omega}=\int\!\!\dx\,\dt\dtb\left(-\frac{1}{2}\bar\Omega_-\Omega_-+\frac{\kappa}{2}\,\bar\Phi_i\Gamma^i_-\Omega_-+{\rm h.c.}\right)\,,
\label{kappa}
\eeq
where $\kappa$ is an additional coupling. It is crucial to note that, since we start from the \nzt model, all fermi multiplets satisfy
\beq
\bar{\mc D}_+\Gamma_-^i=\bar{\mc D}_+\Omega_-=0\,.
\eeq
This constraint turns out to be important, because it guarantees that the interaction term can be recast in half superspace as,
\beq
\frac{\kappa}{2}\int\!\!\dx\,\dt\dtb\,\bar\Phi_i\Gamma^i_-\Omega_-=\frac{\kappa}{2}\int\!\!\dx\,\dt\,\bar{\mc D}_+\bar{\Phi}_i\Gamma^i_-\Omega_-\,.
\eeq  
It was argued in \cite{CS2} that this type of interaction is subject to a ``$D$-term" non-renormalization theorem in $2d$, see also \cite{CCSV}. Therefore, the holomorphic coupling constant $\kappa$ is \emph{not} renormalized. Here we pause and remark that, if one tries to perform the heterotic deformation from \ntt $\mathds{CP}^{N-1}$ GLSM, there would be no non-renormalization theorem to protect the coupling $\kappa$, because in the \ntt case, $\bar{\mc D}_+\Gamma_-^i\propto\Sigma_{(0, 2)}\Phi^i$, see eq.(\ref{fc22}). This differs from the situation in \cite{CCSV}, where the heterotic deformation is indeed performed on \ntt $\mathds{CP}^{N-1}$ NLSM, because the superderivative acting on the fermi multiplet in NLSM automatically vanishes.

Since the coupling $\kappa$ receives no renormalization, we thereby will focus on the $\beta$ function of $\xi$, or say $g_c^{-2}$, in the presence of the coupling constant $\kappa$. Let us first write down the action in components,
\beqn
\mc S_{\Omega}\=\int\!\!\dx\left(i\,\bar\eta_-\partial_{++}\eta_-+\bar HH\right)\nonumber\\[2mm]
\+\kappa\int\!\!\dx\left(i\,\nabla_{++}\bar\phi_i\,\chi^i_-\eta_-+G^i\,\bar\psi_{+\,i}\,\eta_--H\bar{\psi}_{+\,i}\chi^i_-\right)+{\rm h.c.}\,.
\label{kappa2}
\eeqn
The key observation, see also \cite{CCSV}, is that the evolution of the interaction term $i\kappa\nabla_{++}\bar\phi_i\,\chi^i_-\eta_-$ and its 
Hermitian conjugate will give a \emph{finite} shift to the kinetic term of $\phi^i$, i.e.
\beq
\left\langle \kappa\int\!\!\dx\left(i\,\nabla_{++}\bar\phi_i\,\chi^i_-\eta_-\right),\bar \kappa\int\!\!\dy\left(i\,\nabla_{++}\phi^i\,\bar\chi_{-\,i}\bar\eta_-\right)\right\rangle=-\frac{\left|\kappa\right|^2}{4\pi Z_\chi Z_\eta}\int\!\!\dx \left|\nabla_\mu\phi^i\right|^2\,,
\label{shift}
\eeq
where we take fermions as quantum fluctuations and bosons as a background. We write the wave function renormalizations of $\chi_-$ and $\eta_-$ explicitly. It was argued in \cite{CCSV} that this $\left|\kappa\right|^2$ iteration is limited to one-loop in the computation of the quantum correction in the instanton background. Here we have a similar situation -- our $2d$ GLSM admits an (anti-)vortex background, say,
\beq
\nabla_{z}\bar\phi_i=0\,, \orr \nabla_{z}\phi^i=0\,,
\eeq 
where $\nabla_z$ is the Euclidean continuation of $\nabla_{++}$. In this background, the iteration of $\left|\kappa\right|^2$ will not enter higher loops. Nevertheless, the wave function renormalization of the fields $\psi^i_-$ and $\eta_-$ will still enter higher loops evaluation. Therefore, we define a new coupling,
\beq
h^2\equiv\frac{\left|\kappa\right|^2}{Z_\chi Z_\eta}\,,
\eeq 
whose $\beta$ function is given by
\beq
\beta(h^2)=\mu\frac{\partial}{\partial\mu}h^2=h^2(\gamma_\chi+\gamma_\eta)\,,
\label{betah}
\eeq 
where
\beq
\gamma_\chi=-\mu\frac{\partial}{\partial\mu}\log Z_\chi(\mu)\,,\ \ {\rm and}\ \ \gamma_\eta=-\mu\frac{\partial}{\partial\mu}\log Z_\eta(\mu)\,,
\eeq
are the anomalous dimension of the fields $\chi_-^i$ and $\eta_-$\,. 

Now we assemble this addition contribution to the one-loop correction of the holomorphic coupling $\xi$. The imaginary part of eq.(\ref{tc}) is thus modified as
\beq
\frac{2}{g_c^2}=\frac{2}{g^2}-\frac{h^2}{4\pi}+\frac{N}{4\pi}\log Z_\phi(\mu)-\frac{N}{4\pi}\log Z_\chi(\mu)\,.
\eeq
Differentiating with respect to the running scale $\mu$, and using eqs.(\ref{betagamma}) and (\ref{betah}), we arrive at the $\beta$ function for $g_c^2$ in the heterotically deformed \nzt $\mathds{CP}^{N-1}$ GLSM,
\beq
\beta(g_c^2)=-\frac{g_c^4}{4\pi}\frac{N(1+\frac{\gamma_\chi}{2})-h^2(\gamma_\chi+\gamma_\eta)}{1-\frac{N}{8\pi}g_c^2}\,.
\label{hbeta}
\eeq
Finally,  we can to compare eq.(\ref{hbeta}) to the master formula in \cite{CCSV}. In \cite{CCSV}, the kinetic term of the fermion $\chi_-^i$ (in their notation, it was $\psi_R^i$) is non-linearly coupled to the bosonic field $\phi^i$. It makes the definition of the wave function renormalizations of the two theories different up to a scale factor $g_c^2$, i.e.
\beq
Z_{\chi\,\rm here}=g_c^2 Z_{\chi\,\rm there}\,. \eeq
Therefore it leads us to define
\beq
h^{\prime\,2}=h^2 g_c^2\,, \andd \gamma_\chi^\prime=\gamma_\chi+\frac{\beta(g_c^2)}{g_c^2}\,.
\eeq
Under these new definition, we exactly reproduce the master formula in \cite{CCSV},
\beq
\beta(g_c^2)=-\frac{g_c^2}{4\pi}\frac{Ng_c^2(1+\frac{\gamma^\prime_\chi}{2})-h^{\prime\,2}(\gamma^\prime_\chi+\gamma_\eta)}{1-\frac{h^{\prime\,2}}{4\pi}}\,.
\label{hbeta}
\eeq

\vspace{-4mm}

\section*{Acknowledgments}

The research of J.C. is supported in part by the Chinese Academy of Sciences (CAS) Hundred-Talent Program and by Project 11747601 supported by National Natural Science Foundation of China. The work of M.S. is supported in part by DOE grant DE-SC0011842.

\newpage
\begin{appendix}

\section{NLSM of \nzt $\mathds {CP}^{N-1}$ model }
\label{A}

In this appendix, we transform the action of the deformed \nzt $\mathds {CP}^{N-1}$ model of section \ref{5.3} into the corresponding NLSM version. The NLSM can be obtained by integrating out the gauge multiplet of its GLSM cousin at the energy scale $\mu\ll e$. Then, one can study the model in the geometric formalism. First, by integrating the $D$ term, eq.(\ref{D}), one finds the potential
\beq
V_D=\left(\sum_i\bar\phi_i\phi^i-\xi\right)^2\,.
\eeq
On the level of NLSM, it constrains all bosonic fields on $\mathbb S^{2N-1}$, i.e. $\phi^i$ must satisfy the equation
\beq
\sum_i\bar\phi_i\phi^i-\xi=0\,.
\label{c1}
\eeq
On the other hand, integrating the gaugino fields $\lambda_-$ and $\bar\lambda_-$ in eq.(\ref{S1}), we see that the fermion fields $\psi^i_+$ are subject to constraints
\beq
\sum_i\bar\phi_i\psi^i_+=0\,,
\label{c2}
\eeq
implying that $\psi^i_+$'s   live on the tangent bundle of the manifold. In fact, we can rewrite eqs.(\ref{c1}) and (\ref{c2}) together in terms of superfields,
\beq
\sum_i\bar\Phi_i\Phi^i-\xi=0\,.
\label{c3}
\eeq
To obtain the $\mathds{CP}^{N-1}$ model, we need to also take account of the $U(1)$ gauge imposed on $\Phi^i$'s. We can use this gauge to fix one of the chiral multiplet, say the $N$-th field $\Phi^N$, to have its bosonic field \emph{real}\,,
\beq
\Phi^N=\varphi+\sqrt{2}\theta^+\kappa_++\cdots\,,
\eeq
where $\varphi$ now is a real boson, and $\kappa_+$ is its superpartner that is still a complex Weyl fermion. Further we define the gauge invariant coordinates,
\beq
Z^i=z^i+\sqrt{2}\theta^+\zeta_+^i\equiv\frac{\Phi^i}{\Phi^N}\,,\ \ \ {\rm for}\ \ i=1,2,\dots, N-1\,,
\eeq
from which we find
\beq
z^i=\frac{\phi^i}{\varphi}\,,\ \ \ {\rm and }\ \ \zeta_+^i=\frac{1}{\varphi}\left(\psi^i_+-\frac{\phi^i}{\varphi}\kappa_+\right)\,.
\eeq
Now, we can solve for $\Phi^i$ in terms of $Z^i$. From eq.(\ref{c3}), we express $\Phi^N$ as
\beq
\left|\Phi^N\right|^2=\frac{\xi}{1+\bar Z_i Z^i}\,,
\eeq
or, in components,
\beq
\varphi=\frac{\sqrt\xi}{\sqrt{1+\bar z_i z^i}}\equiv\frac{\sqrt\xi}{\rho}\,,\ \ \ {\rm and}\ \ \kappa_+=-\frac{\sqrt\xi}{\rho^3}\,\bar z_i\zeta^i_+\,.
\label{phiN}
\eeq
We then solve 
\beq
\phi^i=\frac{\sqrt\xi}{\rho}z^i\,\ \ \ {\rm and}\ \ \psi^i_+=\frac{\sqrt\xi}{\rho}\left(\delta^i_j-\frac{1}{\rho^2}z^i\bar z_j\right)\zeta^j_+\,, \for i=1, 2, \dots\, N-1\,.
\label{phii}
\eeq
Next, we integrate out the gauge fields $A_\mu$ in eq.(\ref{S1}) and (\ref{S2}), and find  
\beqn
A_{++}\=\frac{i\xi}{2\rho^2}\left(\partial_{++}\bar z_i\,z^i-\bar z_i\partial_{++}z^i\right)+i\,g_{i\bar j}\,\bar\zeta_{+}^{\bar j}\zeta^i_+\,,\nonumber\\[2mm]
A_{--}\=\frac{i\xi}{2\rho^2}\left(\partial_{--}\bar z_i\,z^i-\bar z_i\partial_{--}z^i\right)+i\,\bar\chi_{-\,a}\chi^a_-\,,
\label{Amu}
\eeqn
where, to distinguish the fermi multiplet $\Gamma^a$ from the bosonic one $\Phi^i$, we use the Latin letter ``$a$" to label them, with
$$i=1, 2, \dots, N-1\, \andd a=1, 2, \dots, N\,.$$
Moreover,
\beq
g_{i\bar j}=\frac{\xi}{\rho^2}\left(\delta_{i\bar j}-\frac{1}{\rho^2}\bar z_i z_{\bar j}\right)\,,
\eeq
is the standard Fubini-Study metric on $\mathds {CP}^{N-1}$. The bosonic part of the gauge field is in fact the $U(1)$ piece of the holonomy group $U(N-1)$ of $\mathds{CP}^{N-1}$ \cite{CCSV1}, and couple to the left moving fermion $\chi^a_-$. It implies that the left mover lives on the tautological line bundle $\mc O(-1)$ of $\mathds{CP}^{N-1}$.

Using eqs.(\ref{phiN}), (\ref{phii}) and (\ref{Amu}), we can recast the eqs.(\ref{S1}), (\ref{S2}) and (\ref{fit}) to obtain the NLSM action 
\beqn
\mc S_{\rm NLSM}&=&\int\dx\Large (g_{i\bar j}\,\partial_\mu\bar z^{\bar j}\partial^\mu z^i+ig_{i\bar j}\,\bar\zeta^{\bar j}_+\nabla^{U(N-1)}_{--}\zeta^i_++i\bar\chi_{-\, a}\nabla^{U(1)}_{++}\chi^a_-\nonumber\\[2mm]
&+&
2\left(g_{i\bar j}\,\bar\zeta_+^{\bar j}\zeta_+^i\right)\left(\bar\chi_{-\, a}\chi^a_-\right)\Large),
\label{ztNLSM}
\eeqn
where 
\beqn
&&\nabla^{U(N-1)}\zeta^i_+\equiv d\zeta^i_++\Gamma^i_{jk}dz^j\psi^k_+\,, \with \Gamma^{i}_{jk}=g^{\bar l i}\partial_{k}g_{j\bar l}\,,\nonumber\\[2mm]
&&\nabla^{U(1)}\chi^a_-\equiv d\chi^a_--i\,\omega\,\chi^a_-\,, \with \omega=\frac{i\xi}{2\rho^2}\left(d\bar z_i\,z^i-\bar z_idz^i\right)\,.
\eeqn
One can clearly see that unlike  \ntt $\mathds {CP}^{N-1}$ case, the deformed model has all its left movers living on $\mc O(-1)^{\oplus N}$. We remark here that at the level of NLSM, the study  of isometry/holonomy anomalies is  easy. The $N-1$ right movers $\zeta^i_{+}$ living on tangent bundle of $\mathds{CP}^{N-1}$ contribute to the anomaly proportional to the first Chern class of $T\mathds{CP}^{N-1}$,
\beq
\mc A_{\zeta_+}= c_1(T\mathds{CP}^{N-1})=\frac{N}{4\pi} d\omega\,.
\eeq
On the other hand, the $N$ left movers $\chi^a_{-}$ on $\mc O(-1)^{\oplus N}$ contribute  
\beq
\mc A_{\chi_-}=-\frac{N}{4\pi}d\omega \,.
\eeq
Therefore, the deformed model is anomaly-free as its GLSM cousin, for more details see \cite{CCSV1}.\\

\section{Scaling anomalies: technicalities}
\label{B}

In this Appendix we explain the technique to compute the anomalous Jacobian in section \ref{4.2}, say ${\rm sTr}_{\Phi^i}\mathds 1$ and 
${\rm sTr}_{\Gamma^a_-}\mathds 1$ in eq.(\ref{Jacob}). A careless treatment  of the chiral multiplet $\Phi^i=(\phi^i,\,\psi_+^i)$ seemingly tells us that
\beq
{\rm sTr}_{\Phi^i}\mathds 1=\Tr_{\phi^i}\mathds 1-\Tr_{\psi^i_+}\mathds 1=0\,.
\label{tr1}
\eeq
One has to regularize the above super-trace by introducing regulators. To find a proper regulator, it is sufficient to look at the equation of motion of the
superfield $\Phi^i$ which enters the action $S_{\rm chiral}$, see eq.(\ref{S1}),
\beq
\mathcal D_+\mathcal D_{--}\Phi^i=\cdots\,.
\eeq
We need to further act by $\bar{\mathcal D}_+$ to project the operator equation into the half chiral superspace, i.e.
\beq
\bar{\mathcal D}_+\mathcal D_+\mathcal D_{--}\Phi^i=\bar{\mathcal D}_+\left(\cdots\,\right).
\eeq
After some algebra, we find
\beq
\bar{\mathcal D}_+\mathcal D_+\mathcal D_{--}\Phi^i\propto \left(\nabla_\mu^2+q_i D\right)\phi^i+\sqrt{2}\th\left(\nabla_\mu^2+iq_i B\right)\psi_+^i+\cdots\,.
\eeq
Therefore, the super-trace eq.(\ref{tr1}) is regularized as
\beq
{\rm sTr}_{\Phi^i}\mathds 1=\lim_{M^2\rightarrow\infty}\left(\Tr_{\phi^i} e^{\frac{1}{M^2}\left(\nabla_\mu^2+q_i D\right)}-\Tr_{\psi^i_+} e^{\frac{1}{M^2}\left(\nabla_\mu^2+iq_i B\right)}\right)\,.
\eeq
For trivial fields  $D$ and $B$, the above trace is surely zero. But now let us turn on a non-zero but constant $D$ and $B$ backgrounds. We have
\beqn
\Tr_{\phi^i} e^{\frac{1}{M^2}\left(\nabla_\mu^2+q_i D\right)}\=\int\dx\left<x\left|e^{\frac{\partial_\mu^2}{M^2}}\left(1+\frac{1}{M^2}\left(q_i D+\mathcal O(A_\mu)\right)+\mathcal O\left(\frac{1}{M^4}\right)\right)\right|x\right>\nonumber\\
\=\frac{1}{4\pi}\int\dx\left(M^2+\left(q_i D+\mathcal O(A_\mu)\right)+\mathcal O\left(\frac{1}{M^2}\right)\right), \nonumber\\[2mm]
\Tr_{\psi^i_+} e^{\frac{1}{M^2}\left(\nabla_\mu^2+i q_i B\right)}\=\int\dx\left<x\left|e^{\frac{\partial_\mu^2}{M^2}}\left(1+\frac{1}{M^2}\left(i q_i B+\mathcal O(A_\mu)\right)+\mathcal O\left(\frac{1}{M^4}\right)\right)\right|x\right>\nonumber\\
\=\frac{1}{4\pi}\int\dx\left(M^2+\left(i q_i B+\mathcal O(A_\mu)\right)+\mathcal O\left(\frac{1}{M^2}\right)\right).
\eeqn
Therefore, putting $M^2\rightarrow\infty$, we arrive at
\beq
{\rm sTr}_{\Phi^i}\mathds 1=\frac{q_i}{4\pi}\int\dx\,(D-iB)\,,
\eeq
or, in superspace,
\beq
{\rm sTr}_{\Phi^i}\mathds 1=-i\frac{q_i}{8\pi}\int\dx\,\dt\,\Upsilon_-|_{\thb=0}\,.
\label{19}
\eeq\\
\noindent Similarly, for fermi multiplet $\Gamma^a_-$, we also impose $\bar{\mathcal D}_+\mathcal D_+\mathcal D_{--}$ upon $\Gamma^a_-$ and find,
\beqn
\bar{\mathcal D}_+\mathcal D_+\mathcal D_{--}\Gamma_-^a\=\bar{\mathcal D}_+\mathcal D_{--}\mathcal D_+\Gamma^a_-+\bar{\mathcal D}_+\left(\bar\Upsilon_-\Gamma_-^a\right)\nonumber\\
\!\!&\propto &\!\left(\nabla_\mu^2-i \tilde q_a B\right)\chi^a_--\sqrt{2}\th\left(\nabla_\mu^2-\tilde q_a D\right) G^a+\bar{\mathcal D}_+\left(\bar\Upsilon_-\Gamma_-^a\right)+\cdots
\eeqn
Thus we regularize the super-trace of the fermi multiplet as
\beqn
{\rm sTr}_{\Gamma^a_-}\mathds 1\=\lim_{M^2\rightarrow\infty}\left(-\Tr_{\chi^a_-} e^{\frac{1}{M^2}\left(\nabla_\mu^2-i\tilde q_a B\right)}+\Tr_{G^a} e^{\frac{1}{M^2}\left(\nabla_\mu^2-\tilde q_a D\right)}\right)\nonumber\\[2mm]
\=-\frac{\tilde q_a}{4\pi}\int\dx\,(D-iB)=i\frac{\tilde q_a}{8\pi}\int\dx\,\dt\,\Upsilon_-|_{\thb=0},
\label{21}
\eeqn
cf. section \ref{kon}.
From eq.(\ref{19}) and (\ref{21}), we establish the relation between canonical coupling $\tau_c$ and holomorphic $\tau$ in eq.(\ref{tc}), i.e.
\beq
\tau_c=\tau+\sum_i i\frac{q_i}{4\pi}\log Z_i(\mu)-\sum_a i\frac{\tilde q_a}{4\pi}\log Z_a(\mu)
\eeq
We further remark that, as a consistency check, given a complexified $U(1)$ rotation of the chiral or fermi matter, e.g.
\beq
\Phi^i\longrightarrow e^\alpha\Phi^i\,,
\eeq
the anomalous Jacobian takes the form
\beq
\mathcal J(\alpha)=e^{\alpha\left({\rm sTr}_{\Phi^i}\mathds 1\right)}=e^{\alpha\frac{q_i}{4\pi}\int\dx\,(D-iB)}\,.
\eeq
For real $\alpha$, such as the wave function renormalization or a scale transformation, the anomalous Jacobian only gives a correction to the $D$ term, because $\,{\rm Im}\,\mathcal J(\alpha)$ cancels with the contribution from $\bar\Phi^i$. It simply signals that fermions do not contribute to the one-loop $\beta$-function. On the other hand, for imaginary $\alpha$, it is equivalent to a chiral rotation. We see that $\mathcal J(\alpha)$ and its conjugation only contribute to the flux $B$ term, which gives us the correct chiral anomaly from the chiral fermions $\psi_+^i$ (section \ref{kon}).

\end{appendix}

\end{document}